\documentclass[cameraready]{Interspeech}

\title{Learning from Annotation Uncertainty: Entropy-Aware Curriculum for Speech Emotion Recognition\thanks{To appear in the Proceedings of Interspeech 2026.}}

\author[affiliation={1}, orcid=0009-0009-4951-0929, correspondingauthor]{Zahra}{Omidi}
\author[affiliation={1}, orcid=0000-0003-1382-9929]{John H.L.}{Hansen}

\address{
    $^1$ Center for Robust Speech Systems, \\
    The University of Texas at Dallas, USA 
}

\email{Zahra.Omidi@utdallas.edu, John.Hansen@utdallas.edu}


\usepackage{comment}
\usepackage{amssymb}
\usepackage{multirow}
\usepackage{booktabs}
\usepackage{url}

\begin{document}

\maketitle

\begin{abstract}
Speech emotion recognition (SER) often relies on hard consensus labels that collapse annotator disagreement. We study distribution-based supervision for 9-class SER on MSP-Podcast 2.0 using a WavLM-Base multitask model for categorical emotion and dimensional VAD. Hard-label training is compared with targets from primary and merged primary--secondary annotator vote distributions. Distributional objectives improve alignment with human vote distributions, reducing JSD/KLD relative to hard-label training. Analysis shows that hard supervision partly benefits from assigning ambiguous utterances to the residual \textit{Other} class, whereas distributional supervision redistributes uncertainty across emotion categories. Entropy-stratified evaluation shows that high-ambiguity utterances remain challenging, but distribution-based supervision better captures perceptual uncertainty. These findings support moving beyond hard labels toward targets that reflect listener disagreement.\footnote{\textbf{Code:} \href{https://github.com/zahraomidi/MSP-PODCAST_WavLM}{github.com/zahraomidi/MSP-PODCAST\_WavLM}}
\end{abstract}

\section{Introduction}

Recent advances in speech emotion recognition (SER) leverage large self-supervised encoders (SSEs) such as WavLM, HuBERT, and wav2vec2, yielding substantial gains in robustness. However, most SER systems still assume that each utterance has a single discrete ground-truth label, despite evidence that emotional expressions can support multiple valid interpretations and that annotator disagreement may reflect meaningful perceptual ambiguity rather than noise to be collapsed into a consensus label \cite{barrett2019emotional}.

This limitation is especially relevant for multi-rater speech emotion corpora, where annotations often exhibit disagreement, mixed emotional interpretations, and variability across speakers and contexts \cite{busso2016msp, busso2025msp_podcast, romana2025switchboard, lotfian_msp_2018}. Recent work suggests that such disagreement is not merely annotation noise: annotation variance and label distributions capture structured perceptual information about emotional ambiguity \cite{bnn_ldl_2022, generative_softlabels_2021, learning_annotation_consensus_2025, label_variance_2025}. Likewise, multi-label and mixed-emotion formulations indicate that collapsing annotations into a single consensus label can obscure meaningful variability, as many utterances may be perceived as combinations of prototypical emotion categories rather than strictly discrete classes \cite{chou2024allinclusive, all_inclusive_aggregation_2024, beyond_single_emotion_2025, mixed_emotion_model_2024}.

Despite these advances, results remain difficult to compare due to differences in backbone architectures, label refinement strategies, supervision objectives, and evaluation protocols. In particular, the practical impact of distribution-based supervision in highly imbalanced multi-class benchmarks such as MSP-Podcast remains insufficiently characterized.

In this work, we conduct a controlled study of label-uncertainty supervision within a unified WavLM-Base \cite{chen2022wavlm} framework on the full 9-class MSP-Podcast 2.0 benchmark \cite{busso2025msp_podcast}. We compare hard consensus training with distribution-based supervision derived from primary annotator votes and merged primary--secondary distributions, where the merged distribution combines primary and secondary emotion annotations to capture mixed emotional perceptions. All systems use identical architectures and training protocols within a multitask setting that jointly models categorical emotion and continuous valence, activation, and dominance (VAD). Unlike label refinement approaches that modify or generate new supervisory targets \cite{wen_emolr_2023}, we retain the original annotator distributions to isolate the effect of supervision and objective choice.

We introduce an entropy-aware formulation of SER that treats per-utterance annotation entropy as a fixed dataset property. Entropy is computed from annotator vote distributions and used for ambiguity-stratified evaluation and curriculum design. This allows us to examine how supervision objectives interact with uncertainty and class imbalance, and whether distribution-based training provides measurable benefits over hard labels at different ambiguity levels. We note that entropy estimated from a limited number of annotators is an imperfect proxy for ambiguity; nevertheless, it provides a useful annotation-derived signal for controlled analysis.

\vspace{-1mm}
The main contributions of this work are:
\vspace{-1mm}
\begin{itemize}
\setlength{\itemsep}{0pt}
\setlength{\parskip}{0pt}
\setlength{\parsep}{0pt}
\setlength{\topsep}{0pt}
\item A controlled comparison of hard-label and distribution-based supervision on the 9-class MSP-Podcast 2.0 benchmark.
\item An analysis of primary and merged primary--secondary annotation distributions for modeling ambiguity and non-unique emotion perception in SER.
\item A unified WavLM-based multitask framework that jointly predicts categorical emotion distributions and continuous VAD.
\item An entropy-aware evaluation of SER performance across ambiguity levels, including filtering and weighting curricula on Test1 and Test2.
\end{itemize}

\section{Method}

\subsection{MSP-Podcast 2.0}
Experiments are conducted on MSP-Podcast 2.0, which contains 267,905 utterances from 3,641 speakers. Each utterance is annotated by multiple raters ($\ge5$) with a primary categorical emotion label, optional secondary categorical emotion labels, and dimensional activation, valence, and dominance ratings on a 1--7 Likert scale. We follow the official speaker-independent splits and evaluate on both Test1 and Test2. Split statistics for the official partitions are reported in Table~\ref{tab:data_splits_entropy}.

We retain the full 9-class MSP-Podcast emotion set to preserve the original annotation structure. Collapsing categories would modify the vote distributions and obscure class-dependent ambiguity patterns. All secondary labels are mapped to the same canonical 9-class space before constructing merged primary--secondary distributions.

\subsection{Supervision and Annotation Uncertainty}
To isolate the effect of label uncertainty, we construct all categorical targets from the official MSP-Podcast annotations while keeping the model architecture fixed. 
We compare three supervision regimes. \textbf{Hard consensus supervision} converts each utterance to a one-hot label using the plurality-vote primary emotion.
\textbf{Primary distribution supervision} normalizes primary vote counts into a 9-class probability vector, preserving disagreement among annotators, 
\vspace{-1mm}
\[
P_i=n_i/\sum_j n_j, \quad i=1,\ldots,9,
\]
\textbf{Merged primary--secondary distribution supervision} extends this target by incorporating secondary annotations in the same 9-class space:
\vspace{-1mm}
\[
y_i=\alpha P_i+(1-\alpha)S_i,\quad \alpha\in\{0.9,0.8\},
\]
where $S_i$ is defined analogously from secondary vote counts, and $\alpha\in\{0.9,0.8\}$ gives the 0.9P/0.1S and 0.8P/0.2S targets, respectively. Primary votes capture the dominant perceived emotion, while secondary votes add probability mass to less-dominant perceived emotions.

\textbf{Annotation uncertainty} is computed from the merged primary--secondary distribution and is treated as a fixed utterance-level property. This design keeps the uncertainty signal independent of the training target: a hard-label system, a primary-distribution system, and a merged-distribution system are all assigned the same entropy value for the same utterance. Per-utterance uncertainty is measured using normalized Shannon entropy,
\vspace{-4mm}
\begin{equation}
H_n(p) = - \frac{1}{\log C} \sum_{i=1}^{C} p_i \log p_i,
\end{equation}
\vspace{-3mm}

where $C=9$ and $p$ is the merged annotation distribution. Values lie in $[0,1]$, with larger values indicating broader annotator disagreement or stronger multi-emotion support. Entropy is used only for ambiguity-stratified evaluation and entropy-based curriculum scheduling, not as an additional model output.

\begin{table}[t]
\vspace{-5mm}
\caption{Partition-level statistics of normalized annotation entropy ($H_n$) in MSP-Podcast 2.0.}
\vspace{-3mm}
\centering
\small
\begin{tabular}{lrrrr}
\toprule
Split & Utt. & Median & IQR & \scriptsize{\%($H_n>0.6$)} \\
\midrule
Train & 169{,}190 & 0.437 & 0.252--0.506 & 20.4 \\
Dev   & 34{,}399  & 0.452 & 0.261--0.601 & 25.2 \\
Test1 & 46{,}286  & 0.451 & 0.252--0.605 & 25.9 \\
Test2 & 14{,}822  & 0.481 & 0.287--0.620 & 33.5 \\
\bottomrule
\end{tabular}
\label{tab:data_splits_entropy}
\vspace{-5mm}
\end{table}

\begin{figure}[t]
    \vspace{-4mm}
    \centering
    \includegraphics[width=\linewidth]{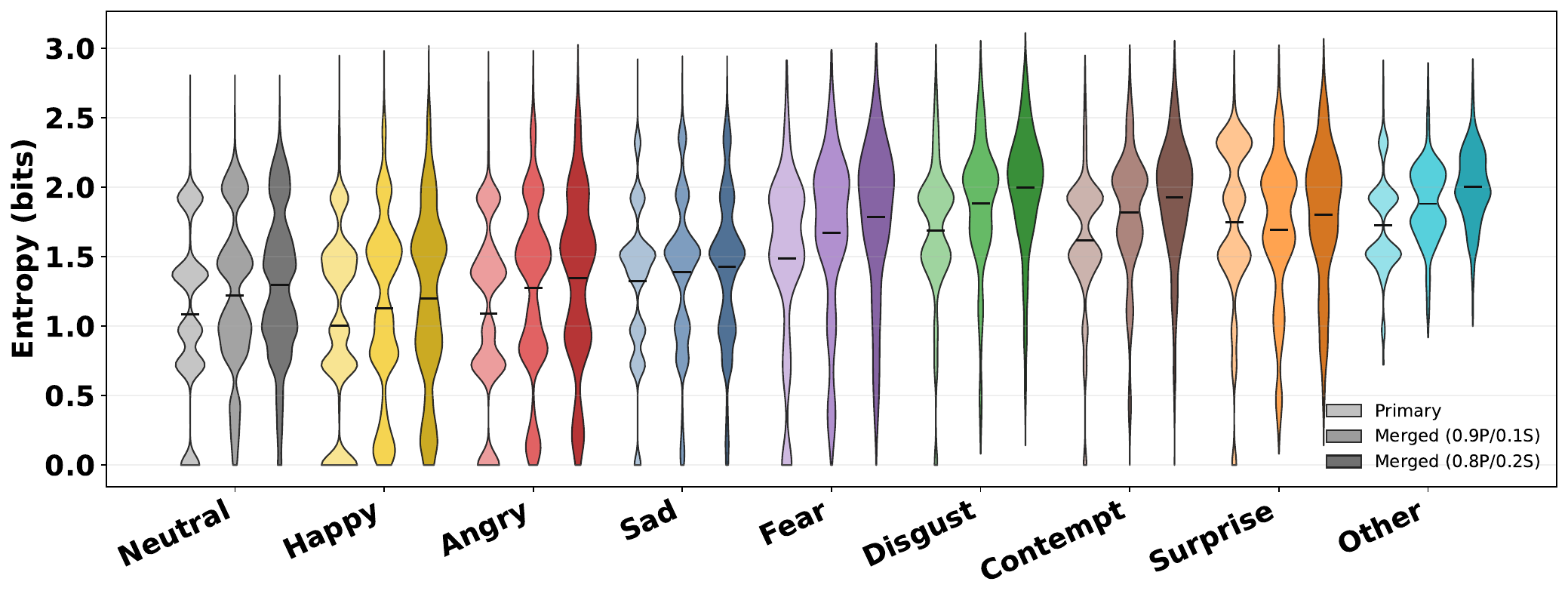}
    \caption{Per-class annotation entropy under primary and merged primary--secondary vote distributions. Merged targets use 0.9P/0.1S and 0.8P/0.2S settings.
    }
    \vspace{-7mm}
    \label{fig:entropy_violin}
\end{figure}

Figure \ref{fig:entropy_violin} shows the entropy structure of the annotation-derived target distributions across emotion classes. Compared with primary vote distributions, both merged primary--secondary settings produce broader entropy profiles, indicating that secondary emotion annotations increase multi-class support in the targets. This effect is class-dependent: fear, disgust, contempt, surprise, and other show higher central tendencies and heavier upper tails, while neutral and happy remain concentrated in lower-entropy regions. The 0.8P/0.2S merge produces greater dispersion than the 0.9P/0.1S merge, reflecting the expected tradeoff between capturing secondary emotional cues and over-smoothing more prototypical classes.

Table~\ref{tab:data_splits_entropy} summarizes normalized entropy across official data partitions. Although median entropy is similar across splits, the evaluation partitions contain a larger proportion of high-entropy utterances, especially Test2. We therefore treat entropy as a fixed annotation-derived property and use it for ambiguity-stratified evaluation and entropy-aware curriculum learning.

\subsection{Model Architecture}

\begin{figure*}[t]
    \centering
    \includegraphics[width=0.87\linewidth]{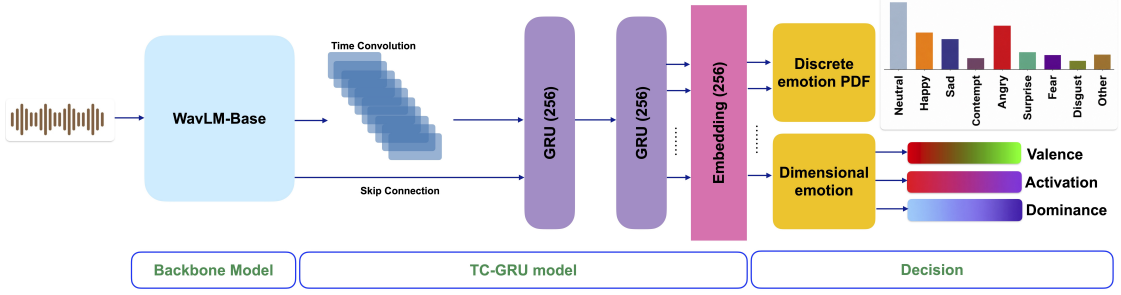}
    \caption{WavLM--TC-GRU multitask architecture with shared 256-dimensional embedding and separate categorical emotion and VAD prediction heads.}
    \label{fig:wavlm-tcgru}
    \vspace{-5mm}
\end{figure*}

Our model builds on the TC-GRU framework of~\cite{mitra2025modeling}, with WavLM-Base replacing the original acoustic backbone and progressive layer unfreezing enabled for fine-tuning. As shown in Fig.~\ref{fig:wavlm-tcgru}, WavLM frame-level representations are processed by a temporal convolution and a two-layer GRU, then projected into a shared 256-dimensional utterance embedding. This shared embedding serves as a compact representation of the utterance-level affective content and is used by two task-specific prediction branches. One branch is a categorical head that predicts the 9-class emotion distribution, while the other is a VAD regression head parameterized by mean and variance to capture uncertainty in continuous affective dimensions. Together, these heads enable joint modeling of discrete emotion categories and continuous affective structure.

WavLM-Base is initialized from pretrained weights and progressively unfrozen during training, with 2, 4, 8, and 12 Transformer layers unfrozen at epochs 1, 2, 4, and 8, respectively. Layer-wise learning-rate scaling is used to stabilize optimization. The complete model contains 95.96M parameters.

\subsection{Training Objective}

The model is trained with a weighted multitask objective combining categorical emotion prediction and VAD regression, with weights 1.0 and 0.3, respectively. The categorical loss depends on the supervision regime: hard-label systems use cross-entropy (CE) or class-balanced cross-entropy (CBCE)~\cite{all_inclusive_aggregation_2024}, while distribution-based systems minimize Kullback--Leibler divergence (KLD) \cite{kullback1951information} between the predicted emotion distribution and the target annotation distribution.

For VAD regression, we use a heteroscedastic Gaussian negative log-likelihood over valence, activation, and dominance, parameterized by predicted mean and variance. A concordance correlation coefficient (CCC) regularization term with weight 0.1 is added to encourage alignment with the continuous ground-truth ratings. When enabled, entropy-based curriculum strategies are applied only to the categorical branch.

\section{Experimental Setup}

\subsection{Optimization}

Models are trained with AdamW using separate learning rates for the WavLM backbone ($1{\times}10^{-5}$) and task-specific heads ($1{\times}10^{-4}$). A NewBob scheduler is applied independently with improvement threshold 0.0025 and annealing factor 0.9. Training uses batch size 32 with mixed precision for up to 18 epochs, and early stopping is based on development Macro-F1 with patience 4. All experiments are conducted on NVIDIA A30 GPUs on the Juno high-performance computing cluster at the University of Texas at Dallas with a fixed random seed.

\subsection{Entropy-Aware Curriculum}

Entropy-based curricula are applied only to the categorical branch. Entropy is computed once from the merged annotation distribution and kept fixed across supervision settings, ensuring that curriculum scheduling is independent of the target type.

We evaluate both standard and reverse curriculum directions. In the \textbf{Filtering} approach, the active training subset is updated according to a predefined quantile schedule over epochs ${1, 2, 4, 8, 12}$ using entropy thresholds corresponding to quantiles ${0.50, 0.60, 0.80, 0.90, 1.00}$. Under the standard curriculum, training begins with lower-entropy utterances and progressively incorporates higher-entropy samples until the full dataset is used. The reverse curriculum follows the complementary schedule, beginning with the full training set and gradually restricting optimization to increasingly higher-entropy regions of the data. In the \textbf{Weighting} approach, all training samples are retained throughout optimization, but each sample's contribution to the categorical cross-entropy loss is scaled by an entropy-dependent weight. Standard weighting assigns larger weights to lower-entropy samples and smaller weights to higher-entropy samples, whereas reverse weighting emphasizes higher-entropy utterances. This design allows us to compare curricula that alter the composition of the training data with those that modify only the effective importance of samples during optimization.

\subsection{Evaluation Protocol}

Categorical performance is reported using Macro-F1, UAR. Distributional alignment is measured using KLD and JSD between predicted and annotation distributions. Dimensional performance is evaluated using CCC for valence, activation, and dominance. For ambiguity-stratified analysis, test samples are grouped by normalized entropy bins to evaluate performance as a function of annotation uncertainty.

\section{Results}

\begin{table*}[t]
\centering
\footnotesize
\caption{Categorical performance and distributional alignment on MSP-Podcast 2.0 Test1/Test2. Target: Hard = hard consensus, Prim = primary distribution, M80/M90 = merged 0.8P/0.2S and 0.9P/0.1S targets. Setting: CE/CBCE = cross-entropy/class-balanced CE, KLD/WKLD = unweighted/weighted KL divergence, Filter/Rev-Filter and Weight/Rev-Weight = standard/reverse entropy curricula. M-F1 and UAR are reported in percent; M-F1, JSD, and KLD include 95\% bootstrap CI half-widths.}
\label{tab:test1_test2_results}
\vspace{-3mm}
\begin{tabular}{llcccc|cccc}
\hline
\textbf{Target} & \textbf{Setting}
& \multicolumn{4}{c|}{\textbf{Test1}}
& \multicolumn{4}{c}{\textbf{Test2}} \\
\cline{3-10}
& & M-F1 & UAR & JSD$\downarrow$ & KLD$\downarrow$ & M-F1 & UAR & JSD$\downarrow$ & KLD$\downarrow$ \\
\hline
Hard & CE & 28.7$\pm$2.1 & 29.3 & .322$\pm$.002 & 1.672$\pm$.016 & 26.3$\pm$2.5 & 21.8 & .340$\pm$.004 & 1.757$\pm$.029 \\
Hard & CBCE & 28.4$\pm$.5 & 27.1 & .329$\pm$.002 & 1.602$\pm$.015 & 20.1$\pm$3.2 & 20.1 & .356$\pm$.003 & 1.704$\pm$.026 \\
\midrule
Prim & KLD & 28.4$\pm$.5 & 28.9 & .203$\pm$.001 & .851$\pm$.006 & 21.0$\pm$3.1 & 22.1 & .211$\pm$.002 & .914$\pm$.009 \\
M80 & KLD & 29.4$\pm$.4 & 32.9 & .235$\pm$.001 & .928$\pm$.005 & 21.6$\pm$.7 & 22.7 & .239$\pm$.002 & .950$\pm$.009 \\
M90 & KLD & 29.2$\pm$.5 & 29.1 & .189$\pm$.001 & .809$\pm$.006 & 28.2$\pm$2.7 & 22.2 & .199$\pm$.002 & .876$\pm$.009 \\
M90 & WKLD & 30.4$\pm$.4 & \textbf{33.1} & .228$\pm$.001 & .919$\pm$.006 & 22.5$\pm$.8 & \textbf{23.1} & .233$\pm$.002 & .940$\pm$.009 \\
\midrule
M90 & Filter & \textbf{34.8$\pm$.5} & 27.9 & \textbf{.185$\pm$.001} & .761$\pm$.005 & 31.5$\pm$3.6 & 22.0 & \textbf{.194$\pm$.002} & .812$\pm$.008 \\
M90 & Rev-Filter & 28.6$\pm$.5 & 28.5 & .215$\pm$.001 & .805$\pm$.004 & 24.1$\pm$4.5 & 21.8 & .213$\pm$.001 & .810$\pm$.007 \\
M90 & Weight & 27.5$\pm$2.1 & 28.0 & .186$\pm$.001 & \textbf{.760$\pm$.005} & \textbf{31.8$\pm$3.3} & 22.2 & .194$\pm$.002 & \textbf{.809$\pm$.008} \\
M90 & Rev-Weight & 28.7$\pm$.5 & 29.1 & .190$\pm$.001 & .788$\pm$.005 & 24.7$\pm$2.6 & 22.8 & .200$\pm$.002 & .845$\pm$.008 \\
\hline
\end{tabular}
\vspace{-5mm}
\end{table*}

Table~\ref{tab:test1_test2_results} summarizes categorical decision performance and distributional alignment across supervision settings. 
Relative to hard CE/CBCE training, all distributional objectives consistently reduce JSD and KLD on both Test1 and Test2, indicating closer agreement between model predictions and annotator vote distributions. Moreover, the narrow bootstrap confidence intervals suggest that these gains are robust under utterance-level resampling.

Macro-F1 differences are smaller than the divergence differences and should be interpreted together with class-level behavior. Since Macro-F1 averages performance across classes, comparable aggregate scores can mask differences in which classes contribute most to the score. In particular, hard-label systems remain competitive in Macro-F1, but part of this performance comes from the residual \textit{Other} category, where Hard--CE obtains Other-class F1 values of 22.8/20.5 and Hard--CBCE obtains 36.0/33.0 on Test1/Test2. This indicates that the strong Macro-F1 of Hard--CBCE is partly driven by its ability to predict the catch-all category, which contains utterances that are difficult to assign to a specific emotion class under consensus labeling.

By contrast, distribution-based systems often produce much lower Other-class F1, such as 0.9/0.9 for Prim--KLD and 1.4/near-zero for M90--KLD. This should not be interpreted as lower Other-class F1 being inherently preferable. Rather, it indicates that the models use the label space differently. Hard-label training treats \textit{Other} as a discrete residual target and can therefore learn it as a separate category. Distributional supervision instead represents ambiguity through probabilities across emotion classes rather than a separate \textit{Other} label.

Entropy-aware curricula further affect this balance. M90--Filter achieves the highest Test1 Macro-F1, suggesting that emphasizing lower-entropy samples early can improve in-split hard-decision performance. M90--Weight achieves the highest Test2 Macro-F1 and the lowest Test2 KLD, indicating better cross-split stability when ambiguous samples remain in training but are weighted by entropy. Reverse curricula, which expose higher-entropy samples earlier, do not outperform the corresponding standard curricula. This suggests that beginning with clearer samples provides a more stable basis for learning categorical structure before introducing stronger annotation ambiguity.

\begin{table}[t]
\centering
\caption{Macro-F1 (\%) stratified by annotation entropy bins on MSP-Podcast 2.0 Test1/Test2. Definitions follow Table~\ref{tab:test1_test2_results}.}
\vspace{-2.5mm}
\label{tab:macroF1_entropy_bins}
\footnotesize
\setlength{\tabcolsep}{4pt}
\renewcommand{\arraystretch}{1.02}
\begin{tabular}{llccc|ccc}
\hline
\textbf{Target} & \textbf{Setting}
& \multicolumn{3}{c|}{\textbf{Test1}}
& \multicolumn{3}{c}{\textbf{Test2}} \\
\cline{3-8}
&
& Low & Mid & High
& Low & Mid & High \\
\hline
Hard & CE     & 32.7 & 24.3 & 14.5 & 21.4 & 19.4 & 14.2 \\
Hard & CBCE   & 29.7 & 24.3 & \textbf{18.1} & 19.6 & 19.4 & 14.8 \\
\midrule
Prim & KLD    & \textbf{34.7} & 25.7 & 12.0 & \textbf{25.7} & 20.7 & 12.0 \\
M80  & KLD    & 31.8 & 27.2 & 17.1 & 22.8 & 21.2 & \textbf{15.3} \\
M90  & KLD    & 34.1 & 26.7 & 17.5 & 25.2 & 21.5 & 14.4 \\
M90  & WKLD   & 33.3 & \textbf{28.0} & \textbf{18.1} & 24.1 & \textbf{22.4} & 14.6 \\
\midrule
M90  & Filter     & 33.6 & 24.1 & 14.9 & 25.0 & 20.4 & 13.3 \\
M90  & Rev-Filter & 34.2 & 25.8 & 16.4 & 24.7 & 21.1 & 13.1 \\
M90  & Weight     & 34.2 & 24.6 & 15.1 & \textbf{25.7} & 20.6 & 13.4 \\
M90  & Rev-Weight & 34.1 & 26.1 & 16.7 & 25.0 & 21.6 & 14.2 \\
\hline
\end{tabular}
\vspace{-3mm}
\end{table}

Table~\ref{tab:macroF1_entropy_bins} reports Macro-F1 after stratifying test utterances by annotation entropy. Across all settings, performance decreases from low- to high-entropy bins, indicating that categorical prediction remains most difficult when annotator disagreement is high.

Distribution-based supervision provides the clearest benefit in the mid-entropy region. On Test1, M90--WKLD achieves the highest mid-entropy Macro-F1 and matches the best high-entropy score, while on Test2 it also yields the strongest mid-entropy performance. These results suggest that distributional targets are most effective when the annotations contain meaningful disagreement without becoming entirely ambiguous.

However, gains in the high-entropy region remain limited. Even the best-performing configurations achieve only modest improvements over hard supervision, and performance remains clustered across methods, indicating that extreme ambiguity continues to challenge categorical recognition.
Thus, uncertainty-aware supervision may help preserve performance under annotation ambiguity, although any gains remain constrained by the limitations of hard categorical evaluation in high-entropy regions. Low Macro-F1 scores in this bin should therefore be interpreted in light of this evaluation mismatch, rather than as a direct indication of the effectiveness of distributional supervision.

\begin{figure}[t!]
    \centering
    \includegraphics[width=\linewidth]{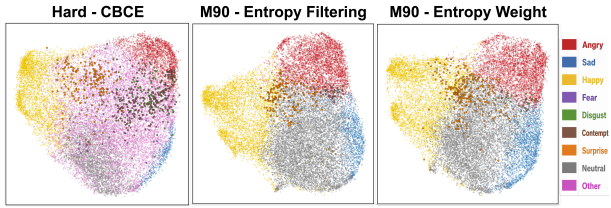}
    \caption{UMAP projection of utterance-level embeddings, colored by predicted emotion.}
    \vspace{-3mm}
    \label{fig:side_by_side_predicted_emotion}
    \vspace{-5mm}
\end{figure}

Figure~\ref{fig:side_by_side_predicted_emotion} qualitatively compares the learned embedding spaces across supervision regimes. Under hard supervision, the projection shows a broad region associated with the residual \textit{Other} class, suggesting that ambiguous utterances are often represented through this catch-all category.

With distribution-based supervision, the embedding space exhibits a more organized relationship among emotion categories. Several classes occupy transitional regions between more prototypical clusters. For example, \textit{disgust} appears between \textit{sad} and \textit{ang}, while \textit{surprise} lies in a region spanning \textit{neu}, \textit{ang}, and \textit{hap}. This structure is consistent with the use of annotator distributions, where an utterance can carry graded support for multiple emotion categories.

Although UMAP visualizations are qualitative, these patterns are consistent with the quantitative results in Table~\ref{tab:test1_test2_results}. Distributional supervision reduces dependence on the residual \textit{Other} category and yields representations that better reflect graded relationships among emotion classes.
 
\section{Discussion}

The results support treating annotation disagreement in SER as structured perceptual information rather than label noise. Since emotion categories can be perceived differently across listeners and contexts~\cite{barrett2019emotional}, hard consensus labels may discard meaningful ambiguity present in multi-rater annotations.

Distribution-based supervision better preserves this ambiguity, consistently reducing JSD and KLD relative to hard-label training. Macro-F1 is less conclusive because it evaluates a single hard decision; hard-label systems partly rely on the residual \textit{Other} class, whereas distributional systems redistribute uncertainty across emotion-specific categories.

Entropy-stratified results show that high-entropy utterances remain challenging under hard categorical evaluation. Distributional supervision and entropy-aware curricula mainly help when disagreement is present but still structured, suggesting that SER evaluation should consider both hard-label performance and distributional fidelity.

\section{Conclusion}

We investigated uncertainty-aware supervision for SSL-based SER on MSP-Podcast 2.0. By treating annotation entropy as an utterance-level property, we analyzed model behavior across different levels of perceptual ambiguity. Training with annotator-derived emotion distributions improves alignment with human vote distributions and reduces reliance on a residual category compared with hard consensus labels.

These findings suggest that hard labels are an incomplete target for subjective emotion annotations. 
Modeling emotion as a distribution over plausible categories provides a more faithful framework for learning from listener disagreement in SER.

\section{Acknowledgments}

The authors acknowledge the High Performance Computing at The University of Texas at Dallas (HPC@UTD) for providing the computing resources and support.

\section{Generative AI Use Disclosure}
During the preparation of this work, the author(s) used Gen AI tools to review and make corrections. After using this tool, the author(s) reviewed and edited the content as needed and take(s) full responsibility for the content of the publication.

\bibliographystyle{IEEEtran}
\bibliography{mybib}

\end{document}